\documentclass[11pt]{article}
\RequirePackage[OT1]{fontenc}
\RequirePackage{amsthm,amsmath}
\RequirePackage[numbers]{natbib}

\numberwithin{equation}{section}
\theoremstyle{plain}
\newtheorem{thm}{Theorem}[section]

\usepackage{graphicx} %
\usepackage{times}                  
\usepackage{tabularx}

\newtheorem{rem}[thm]{Remark}
\newtheorem{cor}[thm]{Corollary}

\newcommand{\barray}{\begin{array}{ll}}   
\newcommand{\earray}{\end{array}}

\usepackage{makeidx}  
\usepackage{amsfonts}
\makeindex            
\usepackage{times}  
\usepackage{graphicx} 
\usepackage{textcomp}

\newcommand{\rr}{{\Bbb R}}

\newcommand{\one}{{\hbox{1{\kern -0.35em}1}}}

\newcommand{\beq}[1]{\begin{eqnarray} \label{#1}}
\newcommand{\eeq}{\end{eqnarray}}
\newcommand{\bed}{\begin{displaymath}}
\newcommand{\eed}{\end{displaymath}}
\newcommand{\bea}{\bed\begin{array}{rl}}
\newcommand{\eea}{\end{array}\eed}
\newcommand{\disp}{\displaystyle}
\newcommand{\al}{\alpha}

\newcommand{\thmref}[1]{Theorem~{\rm \ref{#1}}}

\def\openbox{$\sqcup\llap{$\sqcap$}$}
\def\endproof{\unskip \enskip \null \nobreak \hfill \openbox \par}
\begin{document}

\title{Explicit Solutions for Optimal  Resource Extraction Problems under Regime Switching  L\'{e}vy Models }
\author{Moustapha Pemy\thanks{Department of Mathematics, Towson
University, Towson, MD 21252-0001, mpemy@towson.edu
 } 
 }
\date{}
\maketitle


\begin{abstract} 
This paper studies the problem of optimally extracting nonrenewable natural resources. Taking into account the fact that the market values of the main natural resources i.e. oil, natural gas, copper,...,etc, fluctuate randomly following global and seasonal
 macroeconomic parameters, the prices of natural resources are modeled using Markov switching L\'evy processes. We formulate this optimal extraction problem as an infinite-time horizon optimal control problem. We derive closed-form solutions for the value function as well as the optimal extraction policy. Numerical examples are presented to illustrate these results.  
\end{abstract}




\vspace*{0.2in}

\section{Introduction}

The optimal extraction of nonrenewable natural resources has received a great deal of interest in the literature 
since the early thirties. The first major contribution to this problem was made by  Hotelling \cite{Hotelling}, he proposed an extraction model  in which the commodity price is deterministic and was able to derive an optimal extraction policy. Many economists have 
extended the Hotelling model by taking into account
the uncertainty in the supply and the demand of strategic commodities. Among many others, one can cite the work of  Hanson \cite{hanson, Hanson2}, Solow and Wan \cite{SolowWan}, Pindyck \cite{Pindyck, Pindyck2}, Sweeney \cite{Sweeney},  Lin and  Wagner \cite{LinWagner} for various extensions of the basic Hotelling model.\\
It is self-evident that prices of commodities such as oil, natural gas, copper, and gold are greatly uncertain and fluctuate following divers macroeconomic and global  geopolitical forces.
It is, therefore, crucial to take into account the random dynamic of the commodity value when solving the optimal extraction problem.  In this paper, we use  regime switching L\'{e}vy processes to model natural resources prices. These processes will help us capture both the seasonality and spikes frequently observed in the market prices of natural resources such as oil and natural gas. Moreover, given that the vast majority of mining contracts between mining companies and resource-rich countries are long term contracts, we will study this problem as an infinite time horizon optimal control problem.    
Optimal control problems over finite time and infinite time horizons
 have generated a good deal of interest in the literature, various 
applications have been developed in many areas of science, engineering, and finance. A wide range of techniques have been used to tackle these
 problems. \\
As we all know,  the prices of natural resources such as energy commodities usually feature various spikes and shocks, due to political instabilities in producing countries and the growing global demand for energy. We use L\'{e}vy processes coupled with a hidden Markov chain to capture  jumps and seasonality in commodity prices. L\'{e}vy processes and jump diffusions have also been widely studied in the literature. The optimal control of these processes has
 been investigated by many authors, one can refer to $\O$ksendal and 
Sulem \cite{oksendal}, Hanson \cite{hanson, Hanson2},   and Pemy \cite{ pem3, pem4, pem5}. Roughly speaking, regime switching L\'{e}vy processes consist 
of L\'{e}vy processes with an additional source of randomness, namely,
 a hidden Markov process $(\al(t))_t$ in continuous time or $(\al_n)_n$
 in discrete time. The process $(\al(t))_t$ is a finite states Markov 
chain, it captures the different changes in regime of the L\'{e}vy process.
 Regime switching modeling has been widely used in many fields since
 its introduction by Hamilton \cite{hamilton} in time series analysis.  Many 
authors have studied the control of systems that involve regime switching
 using a hidden Markov chain, one can cite Zhang and Yin \cite{YinZbook}, Pemy and Zhang \cite{PemyZhang}, Pemy \cite{pem, pem2}
 among others. \\ 
In this paper, we treat the problem of finding optimal strategies for extracting a natural resource as a  optimal control problem of Markov switching L\'{e}vy processes in infinite time horizon. The main contribution of this paper is that we fully solve the corresponding Hamilton-Jacobi-Bellman (HJB) equation   which in this case is a system of nonlinear partial integro-differential equations and derive a closed-form representation of the value function and the optimal control strategy.\\ 
The paper is organized as follows. In the next section, we
formulate the problem under consideration.
In Section 3, we solve the HJB equation and we derive both the value function and the optimal extraction policy. And in section 4, we give two numerical examples.

\section{Problem formulation}
Consider a company that has a long term mining lease to extract a strategic natural resource.  Let $m$ be an integer $m\geq2$, and  $\al(t)\in {\cal M} = \{1,2,...,m\}$ be a Markov chain
with generator $Q=(q_{ij})_{m,m}$, i.e., $q_{ij}\geq 0$ for $i\neq j$,  and $\Sigma_{j=1}^m q_{ij}=0$ for $i \in {\cal M}$. In fact, the Markov chain $(\al(t))_t$ will capture various states of the commodity market.
 Let $(\eta_t)_t$ be a L\'{e}vy process, and let $N$ be the Poisson random measure of $(\eta_t)_t$,  for any Borel set $B\subset \rr$, 
$
\disp N(t,B)=\sum_{0<s\leq t}{\bf 1}_B(\eta_s-\eta_{s^-}). 
$
The differential form of $N$ is denoted by $N(dt,dz)$. Let $\nu$ be the L\'{e}vy measure of $(\eta_t)_t$, we have $\nu(B)=E[N(1,B)]$ for any Borel set $B\subset \rr$. 
We define the differential form $\bar{N}(dt,dz)$ as follows,
\bea 
\disp
\bar{N}(dt,dz)=\left\{\begin{array}{ll} N(dt,dz)-\nu(dz)dt \qquad &\hbox{if  } |z|<1\\
   N(dt,dz)&\hbox{if  } |z|\geq 1.   \end{array} \right.
   \eea
From L\'evy-Khintchine formula we have,
\beq{opep}
\disp \int_{\rr}\min(|z|^2,1)\nu(dz)<\infty.
\eeq
Let $X(t)$ denote the price of one unit of a  natural resource  at time $t$. And let $Y(t)$ represent the size of remaining resources at time $t$. 
We assume that the extraction activities for the mining company can be modeled by the process $u(t)$ taking values in a closed and bounded  interval $U$, $u(t)$ is in fact the extraction rate of the resource in question.  The processes $X(t)$ and $Y(t)$ satisfy the following stochastic differential equations. 
\beq{defi3}
\left \{ \begin{array}{ll} \disp\mathrm{d}X(t)= [X(t)\mu(\alpha(t))-\lambda u(t)\big]\mathrm{d}t+\sigma(\alpha(t))X(t)\mathrm{d}W(t)  + \gamma(\alpha(t))X(t)\int_{\rr}z\bar{N}(dt,dz),\\
 dY(t)= -u(t) dt, \\
\disp X(0)=x, \,\,\,\, Y(0) = y \geq 0, \qquad 0\leq t\leq \infty. \end{array} \right.
\eeq   
where $x$ and $y$ are the initial values, $\lambda \in [0,1)$. $W(t)$ is the standard Wiener process on $\rr$,
we assume that $(W(t))_t$, $(\eta_t)_t$ and $(\alpha(t))_t$ are defined on a probability space 
$(\Omega,{\cal F },P)$, and are independent.  The process $u(t)$ is referred as the control process in this model. Moreover, for each $ i \in {\cal M}$, the quantities $\mu(i), \sigma(i)$ and $\gamma(i)$ are assumed to be known constants.

\begin{rem}
Our commodity pricing model (\ref{defi3}) encompasses a wide range of possibilities. Below are some of the particular cases of our general model.
\begin{enumerate}
 \item If the size of the mine is not large enough to influence the price of the commodity then $\lambda =0$. Thus we have the classical exponential L\'{e}vy model for the commodity price 
\beq{ExpoLevy}
\disp\mathrm{d}X(t)= X(t)\bigg(\mu(\alpha(t))\mathrm{d}t+\sigma(\alpha(t))\mathrm{d}W(t)  + \gamma(\alpha(t))\int_{\rr}z\bar{N}(dt,dz)\bigg).
\eeq
This model is appropriate for most mining problems as well as derivative pricing problems.
\item If the size of the mine is large enough or the country where the mine is located is one of the major producers of the commodity in question such as the Saudi Arabia is for oil, then the extraction policies of such a country will definitely affect the world price of the commodity. In this case, we can assume that the drift of the price process will depend on the extraction rate. However, one can foresee  a case where even the diffusion and the jump coefficients are also influenced by the extraction rate. The typical pricing model, in this case, has the form
\beq{ExpoMain}
\disp\mathrm{d}X(t)= (X(t)\mu(\alpha(t))-\lambda u(t))\mathrm{d}t+\sigma(\alpha(t))X(t)\mathrm{d}W(t)  + \gamma(\alpha(t))X(t)\int_{\rr}z\bar{N}(dt,dz),
\eeq    
where $\lambda\in(0,1)$ captures the relative impact of the extracting activities.
\end{enumerate}
In sum, we will study this interesting problem  in its more generalized form as stated in (\ref{defi3}).
\end{rem}
It can be shown that for any Lebesgue measurable  control $u(\cdot)$, the equation (\ref{defi3}) has a unique solution. For more one can refer to  $\O$ksendal and Sulem (2004). For each initial data $(x,y,i)$ we denote by ${\cal U}(x,y,i)$ the set of admissible controls which is just the set of all controls $u(\cdot)$ that are $\{{\cal F}_t\}_{t\geq 0}$-adapted where $ {\cal F}_t = \sigma\{\alpha(\xi),W(\xi), \eta(\xi); \xi\leq t\}$ and such that the equation (\ref{defi3}) has a solution with initial data  $X(0)=x$, $Y(0) =y$, $\al(0) = i$.\\ 
Let $C(u, y)$ be  the extraction cost function, we assume that this function depends on the size of the remaining reserve $Y(t)$ as well as the extraction rate $u(t)$.  Given a discounting factor $r> 0$,  a standard extraction cost function $C(u,y)$ should be increasing in $u$, in particular we assume that $C(u,y)$ is a quadratic function of $u$ and a linear function of $y$. Let $Y(0) = M$ be the size of the initial reserve,  without loss of the generality we will assume that the cost function $C(u,y)$ is given by
\beq{ CostFunction}
C(u, y) =\beta u^2 +\theta u +r\theta y + K,\qquad \beta >0, \,\,\theta>0,\quad K \geq0.
\eeq
 We define the payoff functional as follows
\beq{payoff}
&&J(x,y,i;u)\nonumber \\
&=&E\bigg[\int_0^\infty e^{-r(t)} \Big(X(t)u(t)- C(u(t), Y(t))\Big)dt \bigg{| } X(0)=x,Y(0) = y, \al(0)=i\bigg].
\eeq
 Our goal is to find the control $u^*\in {\cal U}(x,y,i)$  such  that 
\beq{val2}
V(x,y,i)=\sup_{u \in {\cal U} }J(x,y,i;u)=J(x,y,i; u^*).
\eeq
 The function $V(x,y,i)$ is called the value function of the optimal control problem.\\
The process $(X(t),Y(t),\al(t))$ is a Markov process with generator ${\cal L}^w$,   defined as follows
\beq{genra}
\disp
({\cal L}^w f)(x,y,k)&=&\frac{1}{2}\sigma(k)^2x^2\frac{\partial^2f(x,y,k)}{\partial x^2}  \nonumber\\
&&+  (x\mu(k) - \lambda w)\frac{\partial f(x,y,k)}{\partial x}
 +\int_{\rr}\bigg(f(x+\gamma(k)xz,y,k)\nonumber\\
&&-f(x,y,k) -{\bf 1}_{\{|z|<1\}}(z)\frac{\partial f(x,y,k)}{\partial x}\cdot \gamma(k)xz\bigg)\nu(dz) 
\disp  \nonumber\\
&&-  w\frac{\partial f(x,y,k)}{\partial y} +  Qf(x,y,\cdot)(k), 
\eeq
for all  $ x\in \rr, y\in \rr^+, w\in U, k\in{\cal M}, f(\cdot,\cdot,k)\in C^{2,1}_0(\rr\times\rr^+)$
with
\beq{Qgen}
\disp
 Qf(x,y,\cdot)(i)=\sum_{j\not =i}q_{ij}(f(x,y,j)-f(x,y,i)), 
\eeq
the generator of the Markov chain  ($\al_t)_t$.
In order to simplify the notation, we define the operator ${\cal H}$ as follows
\beq{Hamiltonian}
&&{\cal  H} (x,y,i,V(\cdot), V_x(\cdot),V_y(\cdot),V_{xx}(\cdot))  \nonumber\\
&=&rV(x,y,i)- \sup_{u\in U} \Bigg( ({\cal L}^uV)(x,y,i) +(xu-C(u,y)) \Bigg). 
\eeq
It is well known that the value function  $V(x,y,i) $  must formally satisfy the following HJB equation
\beq{sys}
\quad \left \{\begin{array}{ll}
\disp{\cal  H} (x,y,i,V(x,y,i), V_x(x,y,i),V_y(x,y,i),V_{xx}(x,y,i))=0, \\
 \hspace{1 in}  \hbox{for}\,\,(x,y,i)\in \rr\times \rr^+\times {\cal M}. \label{Hamilton}\end{array}\right. 
\eeq
Equation (\ref{sys}) is a system fully nonlinear  of integro-differential equations.

\section{ Closed-form Solutions}
In this section, we show that the optimal extraction strategy is in fact a feedback policy. Using the fact that our running cost functional is a quadratic function of the control variable, we seek a solution of the nonlinear integro-differential equation that is also a quadratic function of the state variables. We have following theorem.    
\begin{thm}\label{MainThm}
The optimal extraction policy is a feedback policy given by 
\beq{optimalcontrol}
u^*(s, i)  \disp =\frac{1}{\beta }\bigg(\frac{1}{2}-\lambda A(i) \bigg)X(s), \qquad s\in[0, \infty),\,\, i\in \{1,...,m\},
\eeq
and the value function is given by 
\beq{valueFunction}
V(x,y,i) = A(i) x^2 -\theta y -  \frac{K}{r},
\eeq
where $A(i), i=1,..., m$, are solutions of the following the system of equations
\beq{needEQ}
&&\disp - r A(i)+ \sigma(i)^2 A(i)  +  \bigg(\mu(i) - \frac{\lambda }{\beta }\big( \frac{1}{2}-\lambda A(i) \big)\bigg) 2A(i) \nonumber\\
& &\disp
 + A(i)\int_{\rr}\bigg((1+\gamma(i)z)^2 -1  -{\bf 1}_{\{|z|<1\}}(z)2 \gamma(i)z\bigg)\nu(dz)
\disp\disp
 +    \sum_{j\ne i} q_{ij}(A(i)-A(j))  \nonumber\\
&& +\frac{1}{\beta }\Big(\frac{1}{4} - \lambda^2 A(i)^2 \Big)=0, \qquad i=1, ..., m.
\eeq
\end{thm}
\paragraph{Proof.}
We will look for a solution of \eqref{sys} in the form
\beq{solfrom}
V(s,x,y,i)  = A(i)x^2+B(i)y + C(i),
\eeq
where for each $i=1,...,m$,  $A(i), B(i), C(i)$,  are real constants. Using the fact that $V(x,y,i)$ must satisfy the equation
\bea
0&\disp= \sup_{u\in U} \Bigg( \frac{1}{2}\sigma(i)^2x^2\frac{\partial^2V(x,y,i)}{\partial x^2}   - rV(x,y,i)\nonumber\\
&\disp+  (x\mu(i) - \lambda u)\frac{\partial V(x,y,i)}{\partial x}
 +\int_{\rr}\bigg(V(x+\gamma(i)xz,y,i)\nonumber\\
&\disp-V(x,y,i) -{\bf 1}_{\{|z|<1\}}(z)\frac{\partial V(x,y,i)}{\partial x}\cdot \gamma(i)xz\bigg)\nu(dz) 
\disp  \nonumber\\
&\disp-  u\frac{\partial V(x,y,i)}{\partial y} +  QV(x,y,\cdot)(i) +(xu-C(u,y)) \Bigg).
\eea
Thus, we have
\beq{yy}
0 &=& \sup_{u\in U} \Bigg{(}\frac{1}{2}\sigma(i)^2x^2 2A(i)  +  (x\mu(i) - \lambda u) 2A(i)x  - r A(i)x^2-r B(i)y- rC(i)\nonumber\\
&&
  - uB(i)+ \int_{\rr}\bigg(A(i)(x+\gamma(i)xz)^2 -A(i)x^2  -{\bf 1}_{\{|z|<1\}}(z)2A(i)x^2 \gamma(i)z\bigg)\nu(dz) 
\disp  \nonumber\\
&&
 +  \sum_{j\ne i} q_{ij}[(A(j)-A(i))x^2 + (B(j)-B(i))y + (C(j)-C(i))] \nonumber\\
&&+ (xu-\beta u^2 - \theta u -r\theta y -K)\Bigg{)}.
\eeq
A necessary condition for optimally in this case is 
\beq{cond}
&&-2\lambda A(i) x  -B(i)+ x-\theta    - 2u\beta  = 0, \nonumber\\
&&u^* = \frac{1}{2\beta }\bigg(-2\lambda A(i) x + x -(\theta+B(i))\bigg).  
\eeq
We set,
\beq{bi}
B(i) = -\theta,\quad \hbox{ for  all}\qquad i= 1, ..., m.
\eeq
Consequently, we should have
\beq{po}
C(i) = -\frac{K}{r},\quad \hbox{ for  all}\qquad i= 1, ..., m.
\eeq
 So the optimum in \eqref{yy} should be attained at $ u^* \disp =\frac{1}{2\beta }\bigg(1-2\lambda A(i) \bigg)x $, thus we should have
\beq{fineEq}
0&=&\disp -r A(i)x^2+ \sigma(i)^2x^2 A(i)  +  \bigg(\mu(i) - \frac{ \lambda}{\beta }\big(\frac{1}{2}-\lambda A(i) \big)\bigg) 2A(i)x^2 \nonumber\\
&& \disp
 + \int_{\rr}\bigg(A(i)x^2(1+\gamma(i)z)^2 -A(i)x^2  -{\bf 1}_{\{|z|<1\}}(z)2A(i)x^2 \gamma(i)z\bigg)\nu(dz) \nonumber
\disp  \nonumber\\
&&\disp
 +    \sum_{j\ne i} q_{ij}(A(j)-A(i))x^2 +\frac{1}{2\beta }\Big(1-2\lambda A(i) \Big)x^2  - \beta \frac{1}{4\beta^2 }\Big(1-2\lambda A(i) \Big)^2x^2\nonumber\\
0&=&\disp - r A(i)+ \sigma(i)^2 A(i)  +  \bigg(\mu(i) - \frac{\lambda }{\beta }\big( \frac{1}{2}-\lambda A(i) \big)\bigg) 2A(i) \nonumber\\
& &\disp
 + A(i)\int_{\rr}\bigg((1+\gamma(i)z)^2 -1  -{\bf 1}_{\{|z|<1\}}(z)2 \gamma(i)z\bigg)\nu(dz)
\disp\disp
 +    \sum_{j\ne i} q_{ij}(A(j)-A(i))  \nonumber\\
&& + \frac{1}{\beta }\Big(1-2\lambda A(i) \Big) \Big(\frac{1}{4 }  +\frac{1}{2 }\lambda A(i) \Big)\Big)\nonumber\\
0&=&\disp - r A(i)+ \sigma(i)^2 A(i)  +  \bigg(\mu(i) - \frac{\lambda }{\beta }\big( \frac{1}{2}-\lambda A(i) \big)\bigg) 2A(i) \nonumber\\
& &\disp
 + A(i)\int_{\rr}\bigg((1+\gamma(i)z)^2 -1  -{\bf 1}_{\{|z|<1\}}(z)2 \gamma(i)z\bigg)\nu(dz)
\disp\disp
 +    \sum_{j\ne i} q_{ij}(A(j)-A(i))  \nonumber\\
&& +\frac{1}{\beta }\Big(\frac{1}{4} - \lambda^2 A(i)^2 \Big), \qquad i=1, ..., m.
\eeq
It is clear that the equations (\ref{fineEq}) are independents of the variable $x$. In fact, we have a system of $m$ nonlinear equations. This system needs to be solved in order to derive the values $A(i), i=1,...m$. This ends the proof of this result. \endproof

\subsection{Optimal extraction strategies when the L\'{e}vy process has finite activity}
In this subsection, we investigate the case where the  L\'{e}vy measure has finite intensity more precisely, the jumps sizes follow an exponential distribution. We assume that the L\'{e}vy measure is of the form
\bea
\nu(dz)  =\left\{\begin{array}{ll} \eta e^{-\eta z} dz & z\geq 0\\
                                             0  & z<0 .
\end{array}
\right .
\eea
  In this particular case, (\ref{needEQ}) becomes 
\beq{need2}
0&=&\disp - r A(i)+ \sigma(i)^2 A(i)  +  \bigg(\mu(i) -  \frac{\lambda}{\beta }\big(\frac{1}{2}-\lambda A(i) \big)\bigg) 2A(i) \nonumber\\
& &\disp
 + A(i)\eta \int_0^\infty(2\gamma(i)z  +\gamma(i)^2z^2)e^{-\eta z} dz  -2\gamma(i)A(i) \eta\int_0^1 z e^{-\eta z}dz
\disp\nonumber\\
&&
 +    \sum_{j\ne i} q_{ij}(A(j)-A(i)) +\frac{1}{\beta }\Big(\frac{1}{4}-\lambda A(i) - \lambda^2 A(i)^2 \Big)\nonumber \\ 
&=&  -r A(i)+ \sigma(i)^2 A(i)  +  \bigg(\mu(i) -  \frac{\lambda}{\beta }\big(\frac{1}{2}-\lambda A(i) \big)\bigg) 2A(i)\nonumber\\
& &\disp
 + 2\gamma(i)A(i)\bigg( 1   -  \frac{1-(1+\eta )e^{-\eta}}{\eta}+\frac{\gamma(i)}{\eta^2}\bigg)
 +    \sum_{j\ne i} q_{ij}(A(j)-A(i))  \nonumber\\
&&+\frac{1}{\beta }\Big(\frac{1}{4} - \lambda^2 A(i)^2 \Big)  \nonumber \\
0&=&\disp\frac{\lambda^2 }{\beta} A(i)^2 + A(i) \Bigg(-r+ \sigma(i)^2 +   2\mu(i) - \frac{\lambda }{\beta } -  \sum_{j\ne i} q_{ij}
 + 2\gamma(i)\frac{\gamma(i)+(1+\eta)\eta e^{-\eta}}{\eta^2}  \Bigg)  \nonumber\\
& &\disp +   \sum_{j\ne i} q_{ij}A(j)+\frac{1}{4\beta}, \quad i=1,...,m. 
\eeq
It is obvious that (\ref{need2}) is  a system of quadratic equations that can be solved in closed form. We have the following corollary.
\begin{cor}
When the L\'{e}vy measure is exponential of the form $\nu(dz) =\eta e^{-\eta z} dz, z>0$ the value function is defined as follows
\bea
V(x,y,i) =A(i) x^2 -\theta y -  \frac{K}{r},
\eea
where the constants $A(i)$ solved the following system of quadratic equations
\beq{Quad}
&&\disp\frac{\lambda^2 }{\beta} A(i)^2 + A(i) \Bigg(-r+ \sigma(i)^2 +   2\mu(i) - \frac{\lambda }{\beta } -  \sum_{j\ne i} q_{ij}
 + 2\gamma(i)\frac{\gamma(i)+(1+\eta)\eta e^{-\eta}}{\eta^2} \Bigg)  \nonumber\\
& &\disp +    \sum_{j\ne i} q_{ij}A(j)+\frac{1}{4\beta}=0, \quad i=1,...,m.
\eeq
\end{cor} 

\subsection{Optimal extraction strategies when the L\'{e}vy process has infinite activity }
It has be shown through empirical evidences that when modeling commodity prices through a L\'{e}vy model, the L\'{e}vy measure has infinite intensity in most cases. For more about this observation one can refer to \cite{AitSahalia2}. It is therefore important that we cover this aspect of the problem and still show that our optimal extraction policy can be derived in closed-form. In that regard, we assume that the L\'{e}vy measure is of the form 
\beq{twoLevy}
\nu(dz) =\left\{\begin{array}{ll}\disp  \frac{e^{-|z|}}{|z|^2} dz & z\neq 0\\
                                             0  & z=0 . 
\end{array}
\right.
\eeq
It is clear that the L\'{e}vy measure defined in \eqref{twoLevy} satisfies the condition \eqref{opep}. From \eqref{needEQ}, we shall have  
\bea
0=&\disp - r A(i)+ \sigma(i)^2 A(i)  +  \bigg(\mu(i) - \frac{\lambda }{\beta }\big( \frac{1}{2}-\lambda A(i) \big)\bigg) 2A(i) \nonumber\\
&\disp
 + A(i)\int_{\rr}\bigg((1+\gamma(i)z)^2 -1  -{\bf 1}_{\{|z|<1\}}(z)2 \gamma(i)z\bigg)\frac{e^{-|z|}}{z^2}dz
\disp\disp
 +    \sum_{j\ne i} q_{ij}(A(i)-A(j))  \nonumber\\
&\disp  +\frac{1}{\beta }\Big(\frac{1}{4} - \lambda^2 A(i)^2 \Big),\\
=&\disp - r A(i)+ \sigma(i)^2 A(i)  +  \bigg(\mu(i) - \frac{\lambda }{\beta }\big( \frac{1}{2}-\lambda A(i) \big)\bigg) 2A(i) \nonumber\\
&\disp
 + A(i)\int_{\rr}\frac{2\gamma(i)z +\gamma(i)^2z^2}{z^2}e^{-|z|} dz -A(i)\int_{-1}^12 \gamma(i)z\frac{e^{-|z|}}{z^2}dz
\disp\disp
 +    \sum_{j\ne i} q_{ij}(A(j)-A(i))  \nonumber\\
&\disp  +\frac{1}{\beta }\Big(\frac{1}{4} - \lambda^2 A(i)^2 \Big),\\ 
=&\disp - r A(i)+ \sigma(i)^2 A(i)  +  \bigg(\mu(i) - \frac{\lambda }{\beta }\big( \frac{1}{2}-\lambda A(i) \big)\bigg) 2A(i) \nonumber\\
&\disp
 + A(i)\int_{-1}^1\frac{2\gamma(i)z}{z^2} e^{-|z|}dz -A(i)\int_{-1}^12 \gamma(i)z\frac{e^{-|z|}}{z^2}dz
\disp\disp
 +    \sum_{j\ne i} q_{ij}(A(j)-A(i))  \nonumber\\
&\disp  +\frac{1}{\beta }\Big(\frac{1}{4} - \lambda^2 A(i)^2 \Big)+ 2A(i)\int_{1}^\infty\frac{\gamma(i)^2z^2}{z^2} e^{-|z|}dz \\ 
&=\disp 
- r A(i)+   \sigma(i)^2 A(i)  +  \bigg(\mu(i) - \frac{\lambda }{\beta }\big( \frac{1}{2}-\lambda A(i) \big)\bigg) 2A(i) \nonumber\\
&\disp
 +    \sum_{j\ne i} q_{ij}(A(j)-A(i))   +\frac{1}{\beta }\Big(\frac{1}{4} - \lambda^2 A(i)^2 \Big)+ 2A(i)\int_{1}^\infty \gamma(i)^2 e^{-z}dz \\
&=\disp 
- r A(i)+   \sigma(i)^2 A(i)  +  \bigg(\mu(i) - \frac{\lambda }{\beta }\big( \frac{1}{2}-\lambda A(i) \big)\bigg) 2A(i) \nonumber\\
&\disp
 +    \sum_{j\ne i} q_{ij}(A(j)-A(i))   +\frac{1}{\beta }\Big(\frac{1}{4} - \lambda^2 A(i)^2 \Big)+ 2A(i) \gamma(i)^2 \frac{1}{e}\\
&=\disp\frac{\lambda^2}{\beta} A(i)^2  + A(i)\bigg(  -r+\sigma(i)^2+2\mu(i)- \frac{\lambda}{\beta} - \sum_{j\neq i} q_{ij} +2 \gamma(i)^2 \bigg)\\
&\disp +\sum_{j\neq i} q_{ij}A(j)+\frac{1}{4\beta}
, \qquad i=1,...,m.
\eea
We therefore have the following corollary which clarifies the expression of  the value function and optimal extraction rate given in \thmref{MainThm} when the jumps activities are infinite and  the L\'{e}vy measure defined as in \eqref{twoLevy}.
\begin{cor}
When the L\'{e}vy measure is of the form $\disp \nu(dz) = \frac{e^{- |z|}}{|z|^2} dz, $ for $z\neq0$, the value function and optimal extraction rate defined in \thmref{MainThm} are such that 
 the constants $A(i), i=1,...,m$, solved the following system of quadratic equations
\beq{Quad2}
\disp 
&&\disp\frac{\lambda^2}{\beta} A(i)^2  + A(i)\bigg(  -r+\sigma(i)^2+2\mu(i)- \frac{\lambda}{\beta} - \sum_{j\neq i} q_{ij} +2 \gamma(i)^2 \bigg)\nonumber \\
&&\disp +\sum_{j\neq i} q_{ij}A(j)+\frac{1}{4\beta}=0
, \qquad i=1,...,m.
\eeq
\end{cor}

\section{Numerical Examples}
\subsection{Example 1: Model with finite  activity}
In this example, we  present the optimal extraction of  an oil field with a known reserve of $Y(0) = M=10$ billion barrels.  We assume that the oil  market has two main movements an uptrend and a downtrend. Thus the Markov chain $(\al_t)_t$ takes two states $ {\cal M}=\{1,2\}$ where $\al(t)=1$ denotes the uptrend and $\al(t)=2$ denotes the downtrend, the yearly discount rate $r=0.02$, the  yearly return vector is $\mu=(0.02,-0.1)$, the  yearly volatility vector is $\sigma=(0.2,0.3)$, the  yearly intensity vector is $\gamma=(0.022,0.03)$,  and the generator of the Markov chain is
\bea 
\disp
Q=\bigg(\begin{array}{ll} -0.3 & 0.3 \\
             0.5 &-0.5 
              \end{array} \bigg).
\eea
The parameter $ \lambda\in[0,1)$ will capture the relative impact of  the oil production on the oil price, in this example $\lambda=0.001$. The extraction cost function is $\disp C(u,y ) = 0.1 u^2 +0.01 u + 0.0002y +10$, so $ \beta =0.1$, $ \theta =0.01$ and $K=10$. The constant $ K$ is seen here as the cost of setting the oil field, which in this case corresponds to \$10 millions. Note that, in the cost function $C(u,y)$, the variable $y$ is in millions and $u$ is in millions per year, and the unit of the cost function $C(u, y)$ is million per year. Assuming that the L\'{e}vy measure is exponential of the form $ \nu(dz)  =e^{-z} dz, z>0$, we have 
\bea
&V(x,y,i) = A(i) x^2 -0.01 y-500,\\
& u^*(s, i)  \disp =5\bigg(1-0.002 A(i)\bigg)X(s), \qquad s\in[0, \infty),\,\, i\in \{1,2\},
\eea
such that $A(1)$ and $A(2)$ solve the system
\bea
\left\{ \begin{array}{ll}  0.00001 A(1)^2-0.281405A(1) +0.3 A(2)+2.5 =0\\
  0.00001 A(2)^2-0.682346A(2) +0.5 A(1)+2.5=0.
\end{array}
\right.
\eea
It is worth noting that the value function $V(x,y,i)$ is given in millions of dollars and the extraction rate $ u^*(s, i)$ is given in millions of barrels per year.
We have the following solutions $A(1) =59.178$ and $A(2) = 47.0599 $. In fact, when we solve the quadratic system  we have four pairs of solutions $( 25706 - 43909 i, 66345 + 34062 i)$, $(4809.48,3732.01 )$, $ (25706 + 43909 i,66345 - 34062i)$  and $ (59.178,  47.0599)$. However, only the pair $ ( 59.178, 47.0599)$ satisfies the constraint that $u^*(s,i)\geq 0, i=1,2$. Therefore, we have
\beq{final}
\left\{\begin{array}{ll} u^*(s,1) =5\bigg(1-0.002 A(1)\bigg)X(s) =4.40822X(s),\\
u^*(s,2) =5\bigg(1-0.002 A(2)\bigg)X(s) =4.5294 X(s),
\end{array}
\right.
\eeq
and \beq{finalVal1}
\left\{\begin{array}{ll} V(x,y,1)=  59.178 x^2 -0.01 y-500,\\
V(x,y,2) =47.0599 x^2 -0.01 y-500.
\end{array}
\right.
\eeq
The equation (\ref{final}) gives the optimal yearly number of millions of barrels we have to extract. These rates are obviously proportional to the oil price $X(s)$, thus our optimal control policy is a feedback policy. One can easily derive the daily optimal extraction by simply dividing the yearly rate by 365. Thus, we have as daily optimal extraction rates
\bea
\left\{\begin{array}{ll} u^*(s,1) =\frac{5}{365}\bigg(1-0.002 A(1)\bigg)X(s) =0.0120773X(s)\quad $ millions of barrels per day$,\\
u^*(s,2) =\frac{5}{365}\bigg(1-0.002 A(2)\bigg)X(s) = 0.0124093X(s)\quad $ millions of barrels per day$.
\end{array}
\right.
\eea
In Figure 1, we represent the value function when the market is up and when the market is down and the L\'{e}vy process has finite jumps activity. 
\subsection{Example 2: Model with infinite activity}
In this example, we repeat the same analysis done in the previous example, the only change is the L\'{e}vy measure. In this case we use the infinity intensity measure $\disp \nu(dz) = \frac{e^{-|z|}}{|z|^2}, z\neq0$. We have the following results. The quadratic system of equations $A(1)$ and $A(2)$ must satisfy is
\bea
\left\{ \begin{array}{ll}  0.00001 A(1)^2-0.249644A(1) +0.3 A(2)+2.5 =0\\
  0.00001 A(2)^2-0.639338A(2) +0.5 A(1)+2.5=0.
\end{array}
\right.
\eea 
The solutions are $( 350.638, 279.35)$,    $(808.633, 642.772)$,     $(24384.8 - 43477.5 i,  63472.7 + 34499.6 i)$    and  $(24384.8 + 43477.5 i,  63472.7 - 34499.6 i)$,  the only solution that satisfies the condition $u^*(s,i)\geq 0$ for all $s\geq 0$ and $i=1,2$ is $A(1) = 350.638, A(2) =   279.35$.  The optimal yearly extraction rates are 
\beq{finalInfinite}
\left\{\begin{array}{ll} u^*(s,1) =5\bigg(1-0.002 A(1)\bigg)X(s) =1.49362X(s),\\
u^*(s,2) =5\bigg(1-0.002 A(2)\bigg)X(s) =2.2065X(s),
\end{array}
\right.
\eeq
and the value function is 
\beq{yuoVal}
\left\{\begin{array}{ll} V(x,y,1)=  350.638 x^2 -0.01 y-500,\\
V(x,y,2) =279.35 x^2 -0.01 y-500.
\end{array}
\right.
\eeq
The optimal daily extraction rates are
\bea
\left\{\begin{array}{ll} u^*(s,1) =\frac{5}{365}\bigg(1-0.002 A(1)\bigg)X(s) =0.00409212X(s)\quad $ millions of barrels per day$,\\
u^*(s,2) =\frac{5}{365}\bigg(1-0.002 A(2)\bigg)X(s) =0.0060452X(s)\quad $ millions of barrels per day$.
\end{array}
\right.
\eea
In Figure 2, we represent the value function when the market is up and when the market is down and the L\'{e}vy process has infinite jumps activity. 
\section{Conclusion}
In this work, we study the optimal natural resource extraction problem when the market value the natural resource follows a regime switching L\'{e}vy process. Our natural resource pricing model captures the main features exhibited by commodities in worldwide exchanges  markets.   We formulate this important economic problem as an optimal control problem and fully solve the nonlinear HJB equation. Our optimal control strategy and the value are derived in closed-form. We end this paper by showing how our result can be easily applied in the optimal management of a massive oil field. 

\begin{center}
\begin{figure}
\includegraphics[height=22cm,width=15cm]{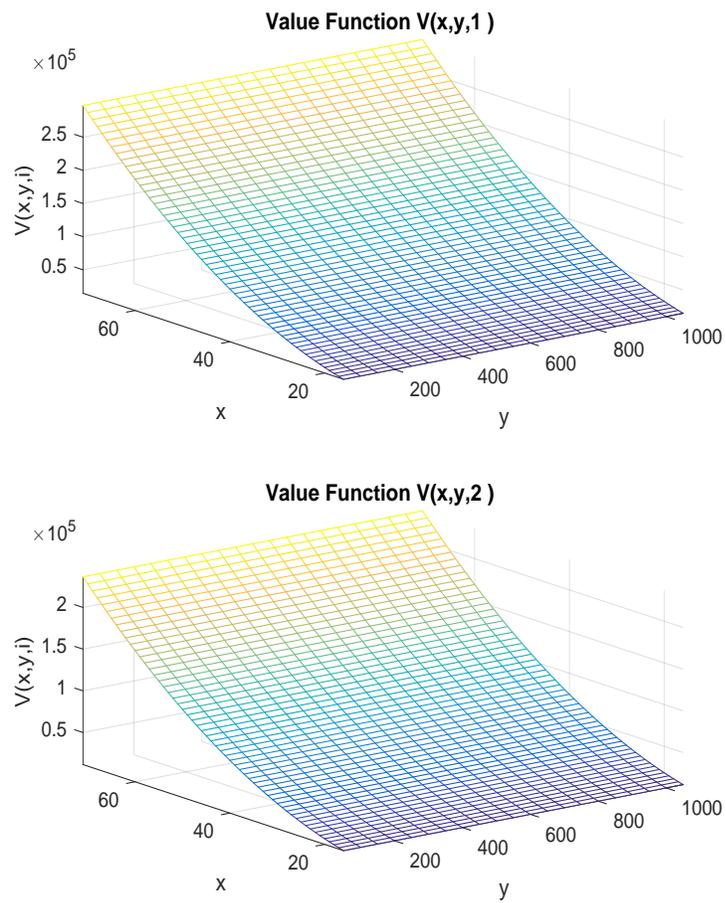}
\caption{Value function when the L\'{e}vy process has finite activity. }
\end{figure}
\end{center}

\begin{center}
\begin{figure}
\includegraphics[height=22cm,width=15cm]{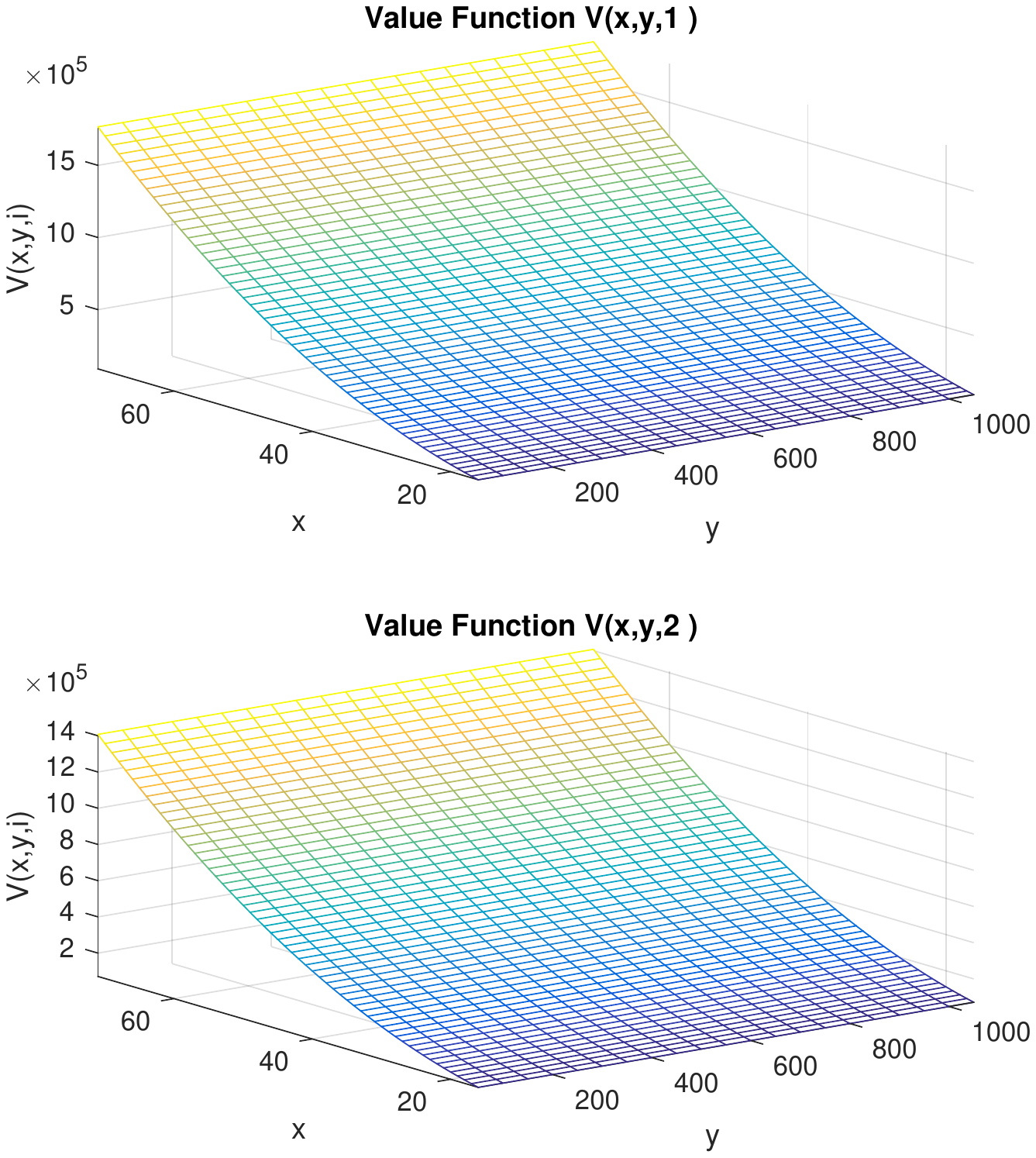}
\caption{Value function when the L\'{e}vy process has infinite activity . }
\end{figure}
\end{center}

\end{document}